\documentclass[9pt,twocolumn,twoside]{osajnl}
\journal{ol} 
\setboolean{shortarticle}{true}

\usepackage{lineno}
\usepackage{subfigure}

\title{Force Detection Sensitivity Spectrum Calibration of Levitated Nanomechanical Sensor Using Harmonic Coulomb Force}

\author[1]{Zhenhai Fu}
\author[1]{shaochong zhu}
\author[1]{ying dong}
\author[1,2]{xingfan chen}
\author[1,2]{Huizhu Hu}
\author[1,*]{xiaowen gao}

\affil[1]{Quantum Sensing Center, Zhejiang Lab, Hangzhou, 310000 China}
\affil[2]{State Key Laboratory of Modern Optical Instruments, College of Optical Science and Engineering, Zhejiang University, Hangzhou, 310027, China}

\affil[*]{Corresponding author: gaoxw@zhejianglab.com}

\begin{abstract}
	Oscillators based on levitated particles are promising for the development of ultrasensitive force detectors. 
	The theoretical performance of levitated nanomechanical sensors is usually characterized by the so-called thermal noise limit force detection sensitivity, which does not exhibit spectral specificity in practical measurements.
	To characterize the actual detection performance, we propose a method for the force detection sensitivity calibration of a levitated nanomechanical sensor based on the harmonic Coulomb force. 
	Utilizing the measured transfer function, we obtained the force detection sensitivity spectrum from the position spectrum.
	Although the thermal noise limit force detection sensitivity of the system reached $\rm\left( {4.39 \pm 0.62} \right) \times {10^{ - 20}} N/H{z^{1/2}}$ at $\rm{2.4\times10^{-6} mbar}$ with feedback cooling, the measured sensitivity away from the resonance was of the order of $\rm10^{-17} N/Hz^{1/2}$ based on the existing detection noise level.
	The calibration method established in our study is applicable to the performance evaluation of any optical levitation system for high-sensitivity force measurements.
\end{abstract}

\setboolean{displaycopyright}{true}

\begin{document}
	
	\maketitle
	
	\section{Introduction}
	Optical trapping and manipulation of nanoparticles in a high vacuum has great potential application to force detectors for fundamental forces and interactions, providing an approach for testing short-range gravitational physics \cite{rider2016search, moore2014search}, evaluation of surface forces, including the Casimir effect \cite{ether2015probing}, and detection of high-frequency gravitational waves \cite{arvanitaki2013detecting}. When isolated from physical contact with the environment, the thermally limited detection sensitivity of levitated oscillators increases with increasing vacuum, the tunable resonance frequency of which ranges from tens/hundreds of Hz for microparticles to tens of kHz for nanoparticles \cite{moore2021searching}. For instance, under an ultrahigh vacuum of $ 10^{-8} $mbar and at room temperature, a Q value of up to $ 10^{10} $ and sensitivity of $\rm{10^{-20} N/Hz^{1/2}}$ can be achieved \cite{millen2020optomechanics}. 
	
	Recent studies have demonstrated levitated oscillators in a vacuum with zeptonewton-level force sensitivities \cite{ranjit2016zeptonewton} and capability of probing and calibrating electrostatic force and gravity \cite{hempston2017force, hebestreit2018calibration, hebestreit2018sensing}. Nevertheless, thermally limited detection sensitivity fails to accurately describe the actual performance of the system under a high vacuum, where thermal noise is no longer dominant but suppressed by irrelevant noises, such as those from light and the detector \cite{millen2020optomechanics}. The force detection sensitivity spectrum (FDSS) calculated from the transfer function of the oscillator represents the performance of the system, as it considers all the noise sources mentioned above \cite{kawasaki2020high}. Furthermore, as such noise is distinct from white noise due to its differences across the spectrum, the calibrated FDSS also exhibits spectral specificity. 
	
	Here, we propose and demonstrate a novel technique utilizing Coulomb force to calibrate the FDSS of an optically levitated nanoparticle oscillator. Transfer function measurement was first achieved at different pressures and compared with the theoretical curves including crosstalk between three orthogonal modes. Then, we obtained the FDSS at different pressures based on the measured displacement power spectral density (PSD). Finally, feedback cooling was applied in a high vacuum to obtain the ultimate detection sensitivity of the system. The calibration method established in our study is applicable to the performance evaluation of any optical levitation system for high-sensitivity force measurements.
	
	\section{Theoretical principles}
	
	In this section, we describe the theoretical principles of calibrating the FDSS of a levitated nanomechanical sensor. The levitated nanomechanical resonator converts the external force  $F(t)$ into displacement of the resonator $ q(t) $, where $ q\in{ \left\lbrace x,y,z \right\rbrace } $  represents mutually orthogonal axes. 
	The response of system generated by inherent fluctuating force $F_{\mathrm{th}}(t)=\sqrt{2k_BTm\Gamma}\eta(t) $  can be expressed in terms of power spectral density as follows
	\begin{equation}
		S_{th}^q(\omega ) = |\widetilde{\chi}(\omega){|^2}S_{th}^F(\omega ) + \zeta
		\label{noise} 
	\end{equation}
	
	where $ \widetilde{\chi}(\omega) $ represents the transfer function of the system. 
	Here, we assume that the measurement noise $ \zeta $ is independent of the force sensing process and mainly originates from light sources, detection devices, environmental vibration, etc.
	Therefore, when the minimum detectable force $ {F_{\min }}(t) $ is added, the response can be rewritten as follows
	\begin{equation}
		S_{\min }^q(\omega ) = |\widetilde{\chi}(\omega){|^2}\left[ {S_{th}^F(\omega ) + S_{\min }^F(\omega )} \right] + \zeta
		\label{signal} 
	\end{equation}
	
	When we characterize the force detection performance of the resonator, the response generated by  inherent fluctuating force can be regarded as the noise.
	The FDSS of the system is equivalent to the power spectrum density of the minimum detectable force, and can be obtained by eliminating the measurement noise term from Eq.\ref{noise}and \ref{signal} for the case of $ {\rm{SNR}}=1 $ 
	\begin{equation}
		S_{FDSS }(\omega ) = \sqrt {\frac{{S_{th}^q(\omega )}}{{|\tilde \chi (\omega ){|^2}}}} 
		\label{eq-FDSS}
	\end{equation}
	
	Therefore, the force detection sensitivity spectrum can be deduced from the measured transfer function and the displacement power spectral density without external force.
	When thermal noise is dominant in the system noise and the noise term $  \zeta $ can be ignored in Eq.\ref{noise}, FDSS can be approximated by the theoretically thermal-noise-limit detection sensitivity, which remain frequency independent as follows:
	\begin{equation}
		{S_{FDSS}}(\omega ) \simeq S_{\min }^{F_{th}}(\omega )  {\rm{ = }}\sqrt {2{k_B}Tm\Gamma } 
		\label{det-sen}
	\end{equation}
	
	where $ k_B $ is Boltzmann’s constantb, $ T $ is temperature of the center of mass (COM), $ \Gamma $ is damping rate.
	Applying an external force $  F_{ext}(t)  $ with known amplitude at a frequency of $ \omega_{ext} $ and obtaining the corresponding response ${q_{ext}}\left( t \right)$, the transfer function can be obtained according to the following formula
	\begin{equation}
		\tilde \chi ({\omega _{ext}}) = \frac{{{{\tilde q}_{ext}}({\omega _{{\rm{ext}}}})}}{{{{\tilde F}_{ext}}\left( {{\omega _{{\rm{ext}}}}} \right)}}
	\end{equation}
	
	where $ {{{\tilde q}_{ext}}({\omega _{{\rm{ext}}}})} $ and $ {{{\tilde F}_{ext}}\left( {{\omega _{{\rm{ext}}}}} \right)} $ are the Fourier transforms of $  F_{\omega_{ext}}(t)  $ and ${q_{ext}}\left( t \right)$, respectively.
	Experimentally, a harmonic electric field of $ \mathbf{E}\left(t\right)=E_0cos{\left(\omega_{ext}t\right)}\mathbf{x} $  with a frequency of $ \omega_{ext} $ was applied to the nanoparticle along the x-axis, exerting a harmonic driving force of $ \mathbf{F}_{ext}\left(t\right)=qE_0cos{\left(\omega_{ext}t\right)}\mathbf{x} $  onto the nanoparticle, where $ q=Nq_e $ is the nanoparticle charge. 
	Response component $ {\widetilde{q}}_{ext}(\omega_{\mathrm{ext}}) $  of the system can be obtained by deducting thermal noise component from total response.

	\section{Experimental setup}
\begin{figure}[htbp]
		\centering{\includegraphics[width=0.8\linewidth]{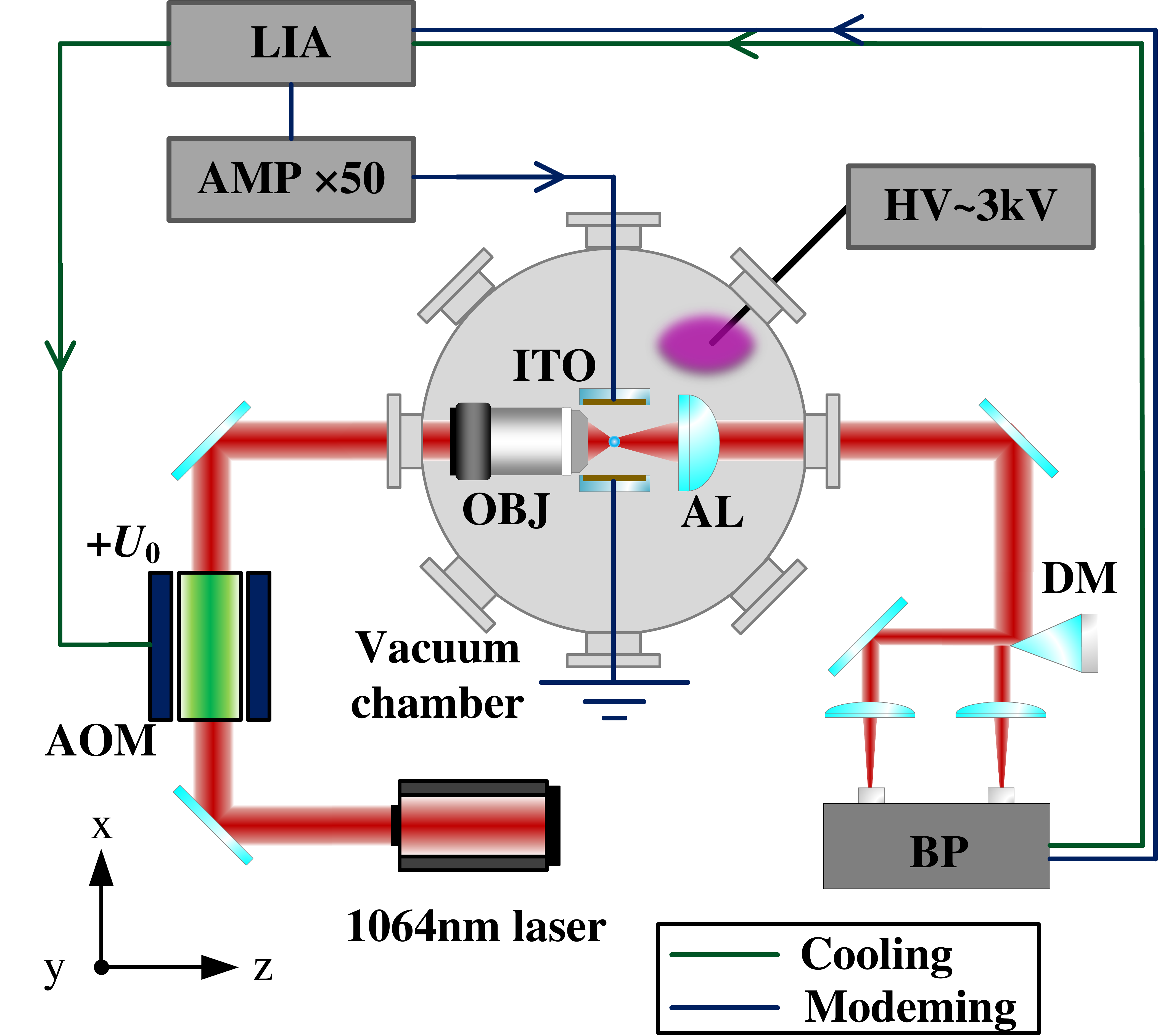}}
		\caption{Schematic of the experimental setup. 
		A single-beam optical trap was formed by focusing a 1064-nm Gaussian beam with a high NA microscope objective (OBJ). The motion of the nanoparticle along the horizontal x-axis was detected by a differential detection scheme consisting of an aspheric lens (AL), D-shaped mirror(DM), and a balanced photodetector (BP). The amplified harmonic signal generated by the lock-in amplifier (LIA) was loaded onto the ITO electrode as the electric driving signal. Two bare wires connecting the high DC power supply outside the vacuum chamber were used to manipulate the charge of the nanosphere through corona discharge. The lock-in amplifier was used to extract the electric drive response component, as shown by the green lines, and to cool the COM of the nanosphere, as shown by the blue lines.
	}
		\label{setup1}
	\end{figure}

	As illustrated in Fig.\ref{setup1}, a Gaussian beam generated from a 1064-nm laser (ALS-IR-1064) was converged by a high numerical aperture (NA) microscope objective of 0.8 (Nikon TU plan ELWD) to trap a silica nanoparticle at its focus. Light scattered from the nanoparticle was collimated by an aspheric lens, split by a horizontally-placed D-shaped mirror, and then collected by a balanced photodetector, from which the motion signal on the x-axis was deduced. A pair of indium tin oxide (ITO) electrodes were coated onto two slides of 1×5 cm quartz glass, placed 10 mm apart perpendicular to the beam, which produced a uniform electric field surrounding the optical trap \cite{park2015parametric}. 
	Based on the ITO electrode geometry and applied voltage of $U=50 V$,  the electric field  $ {E_0} = 3800 \pm 10 {\rm{V/m}} $ was calculated using finite element analysis. 
	The harmonic driving signal of the electrodes was generated by amplifying the original signal using a lock-in amplifier (Zurich Instruments MFLI) with a 50× power amplification up to 300 Vpp. 
	Bare wires were mounted approximately 1 cm from the center of the trap, importing a high DC voltage of up to 3 kV from outside the chamber. Through corona discharge, this high voltage created a plasma of both positive and negative ions to manipulate the nanoparticle charge\cite{lieberman2005principles, frimmer2017controlling, ricci2019accurate}. The lock-in amplifier processed the nanoparticle motion signal in two distinct modes. Under the sweep mode, a ratio of $ \widetilde{\chi}(\omega_{\mathrm{ext}})={\widetilde{q}}_{ext}(\omega_{\mathrm{ext}})/{\widetilde{F}}_{ext}\left(\omega_{\mathrm{ext}}\right)  $ was calculated over a wide frequency range from 1 to 500 kHz, which covers twice the resonant frequency of the x-mode; in the proportional–integral–derivative mode, a frequency-doubled signal was loaded on the acousto-optic modulator to cool the COM of the nanosphere in a high vacuum.
	\section{Results and discussion}
	\subsection{Calibration of basic parameters}
\begin{figure}[htbp]
		\centering {\includegraphics[width=\linewidth]{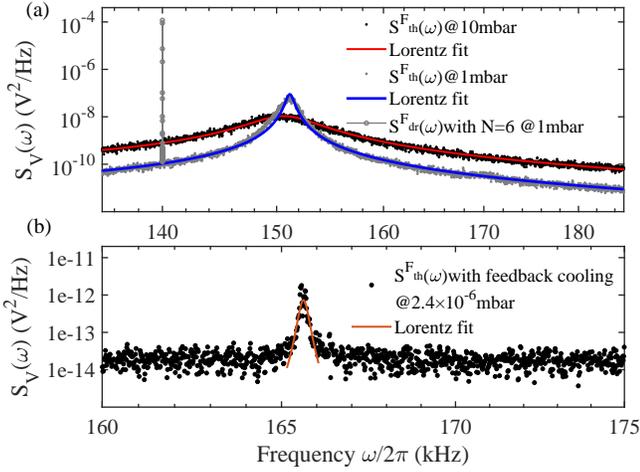}}
		\caption{PSD and Lorentz fitting. 
			(a) 10 mbar and 1 mbar without cooling.  The resonance frequency was approximately $ 151.1 {\rm{kHz}} $. The gray line represents the electric driving response with a charge number of N=6 at 140 kHz.
			(b) $2.4\times10^{-6}$ mbar with feedback cooling. A resonance frequency of $ 165.6 {\rm{kHz}} $ and fitting damping rate of ${\Gamma _{cool}}/2\pi  = 144.8 \pm 18.5 {\rm{Hz}}$ were extracted by Lorentz fitting.
		}
		\label{psd}
	\end{figure}

	We first calibrated the basic parameters, including the calibration factor $ c_{uq} $  between the signal voltage $ U\left(t\right) $and particle position $ q\left(t\right) $, particle size $r$, and the step value of a single charge $ \Delta U_{qe} $. As shown in Fig.\ref{psd}(a), a calibration factor of  $ c_{uq}=\rm{ 80.7 \pm 0.5 mV/nm }$ in the x-axis was obtained at a pressure of 10 mbar, where the particle and environment were in thermal equilibrium \cite{hebestreit2018calibration}. 
	The resonance frequency and damping rate were extracted by Lorentz fitting. The radius $ r=\rm {80.8\pm3.1 nm} $ of the particle was deduced from a fitting damping rate of $\Gamma_x /2\pi={\rm{ 8.2 \pm 0.4 kHz }}$ based on the principle of gas dynamics \cite{beresnev1990motion}. 
	We further reduced the pressure to $\rm2.4\times10^{-6} mbar$ and cooled the COM of the nanosphere to $ {T_{cool}}{\rm{ = 15}}{\rm{.80}} \pm {\rm{3}}{\rm{.58 mK}}$ by applying parametric feedback cooling \cite{gieseler2012subkelvin}, as shown in Fig\ref{psd}(b). A thermal noise limit force detection sensitivity of $\rm\left( {4.39 \pm 0.62} \right) \times {10^{ - 20}} N/H{z^{1/2}}$ is possible according to Eq.\ref{det-sen}. 
	
\begin{figure}[htbp]
		\centering {\includegraphics[width=\linewidth]{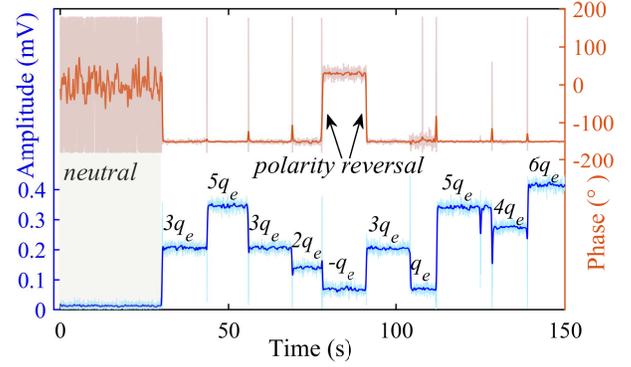}}
		\caption{Charge control at 1 mbar. A driving signal with an amplitude of 50 Vpk and frequency of 140 kHz was applied to the ITO electrode.  The lock-in amplifier demodulated the amplitude (blue) and phase (red) at the driving frequency with a sampling rate of 1 kHz. The dark colored curves are 1 s smoothed data from the original data (light color) to show a more distinguishable step. The amplitude changed after each discharge, and the corresponding charge quantity is marked at each step.
		}
		\label{charge1}
	\end{figure}

	The charge of nanoparticle was then controlled by corona discharge at 1mbar and Fig.\ref{charge1} shows the amplitude and phase that LIA demodulated from the motion signal in x-axis. 
	The particles initially showed a non-null net charge of the order of ten elementary charges and reached an electrically neutral state after several discharge processes.
	In the electrically neutral state, the phase demodulated from the motion signal at the driving frequency is random, and the amplitude is close to the thermal noise level.
	Particle's polarity could be reversed with a 180° shift in phase during the discharge process.
	Then, a minimum amplitude step of $69.46 \pm 1.35\mu  {\rm{V}}$ was obtained by multiple discharges, corresponding to the change of a single charge. Using this information from a single charge step, we calculated the precise charge in real time and then adjusted the nanoparticle to a suitable charge state. We found that when the air pressure dropped below 0.1 mbar, the charge of the nanosphere was prone to changing abruptly. Although we could not specify the mechanism underlying this phenomenon, we calculated the charge at the beginning and end of each scanning to ensure that it remained constant as a known parameter throughout the entire process. 
	\subsection{Calibration of transfer functions}
\begin{figure}[htbp]
		\centering {\includegraphics[width=\linewidth]{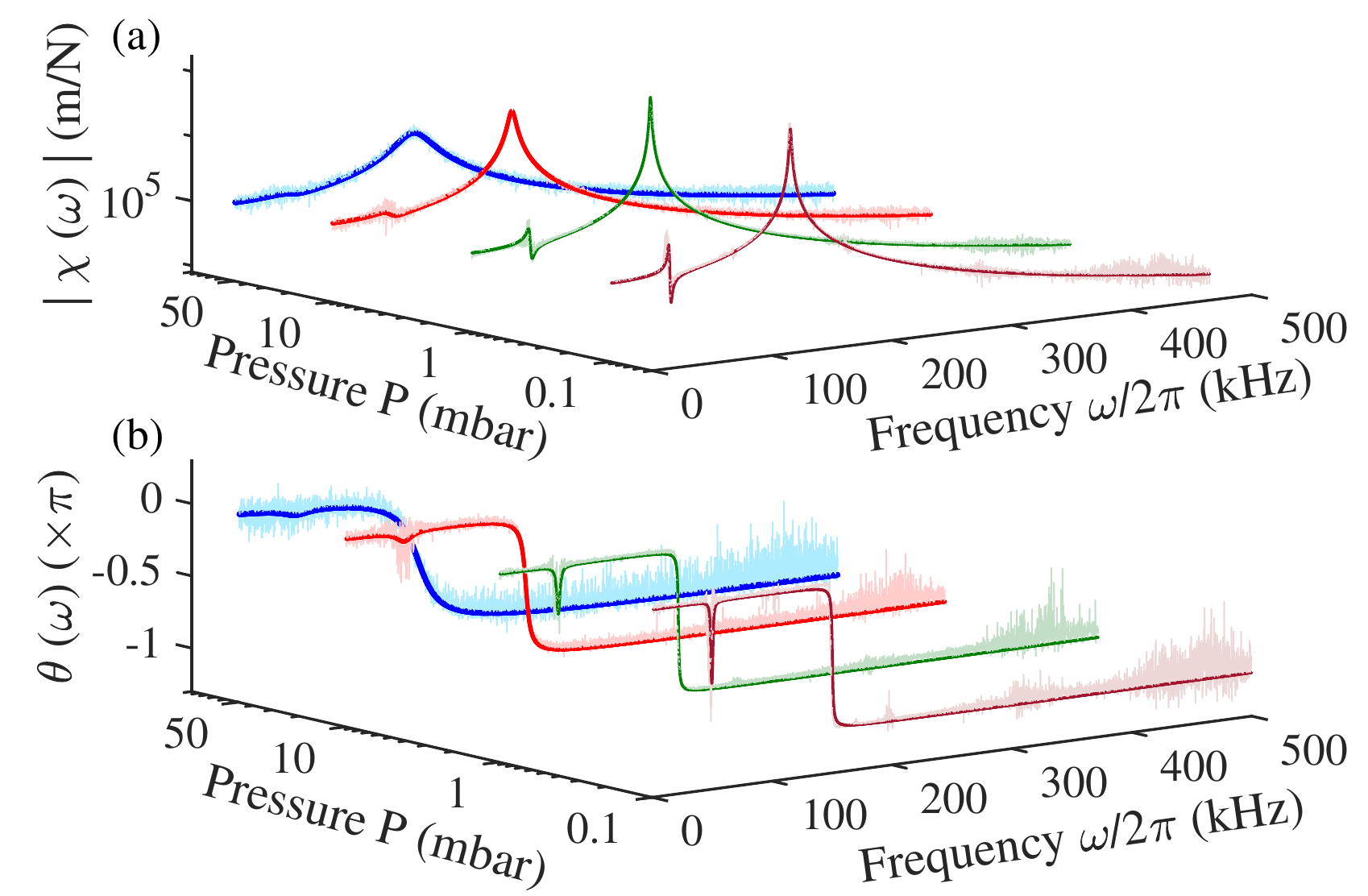}}
		\caption{Transfer functions of a typical nanomechanical resonator at pressures of 50, 10, 1, and 0.1 mbar, where the  (a) amplitude and (b) phase of the complex are shown. Data measured using the method described above are shown in light colors, and theoretical curves involving crosstalk modes are shown in darker colors.}
		\label{tf}
	\end{figure}

	We measured the transfer function at different pressures and compared it with the theoretically fitted curve, as shown in Fig.\ref{tf}.
	The linear transfer function of a levitated nanomechanical resonator can be derived from the Langevin equation \cite{langevin1908theory, lemons1997paul} as follows:
		\begin{equation}
			\widetilde{\chi}(\omega)=\frac{1}{m\left(\omega_0^2-\omega^2+i\Gamma\omega\right)}
			\label{eq-TF}
		\end{equation}
	Due to the cross talk inherent to the detection scheme, the transfer function calibrated from the single-axis signal did not behave as a single mode, but instead included crosstalk modes from other orthogonal axes expressed as follows:  
	\begin{equation}
		{\tilde \chi _x}(\omega ) = {\tilde \chi _{xx}}(\omega ) + \sum\limits_{i = y,z} {{\varepsilon _{x - i}}{{\tilde \chi }_{x - i}}(\omega )} 
		\label{ct-TF}
	\end{equation}
	where $ {\widetilde{\chi}}_{x-i}(\omega),i\in\left(x,y\right) $ represents the crosstalk modes, and $ \varepsilon_{x-i} $ is the corresponding coupling coefficient. Each crosstalk can be written in the form of Eq.\ref{eq-TF} based on its resonant frequency and damping rate.
	In the experiment, the y-mode crosstalk could be eliminated by rotating the direction of the D-shaped mirror and surface of the detector \cite{zhu2020displacement}. However, due to different detection strategies, the z-mode crosstalk of $ \omega_{xz}=48.5 \mathrm{kHz} $ remained with a coupling coefficient of $ \varepsilon_{x-z}\approx0.008 $. 
	The crosstalk of the z-mode accounted for the extra tongue in the amplitude response and valley in the phase response, which became sharper with a decrease in pressure.
	
	The measured transfer function was consistent with the theoretical curve at different pressures, indicating that the mechanical response of the nanosphere resonator was linear. In fact, we could deduce from the amplitude of the electric field force $ F_0=3.64 \mathrm{fN} $ and optical trap stiffness $ k_{xx}=3.9\mu \rm{ N/m} $ that the displacement of the nanosphere due to the electrical force was less than 1 nm, suggesting that it was constrained in the linear range of the trap.
	
	\subsection{Calibration of FDSS}
\begin{figure}[htbp]
		\centering {\includegraphics[width=\linewidth]{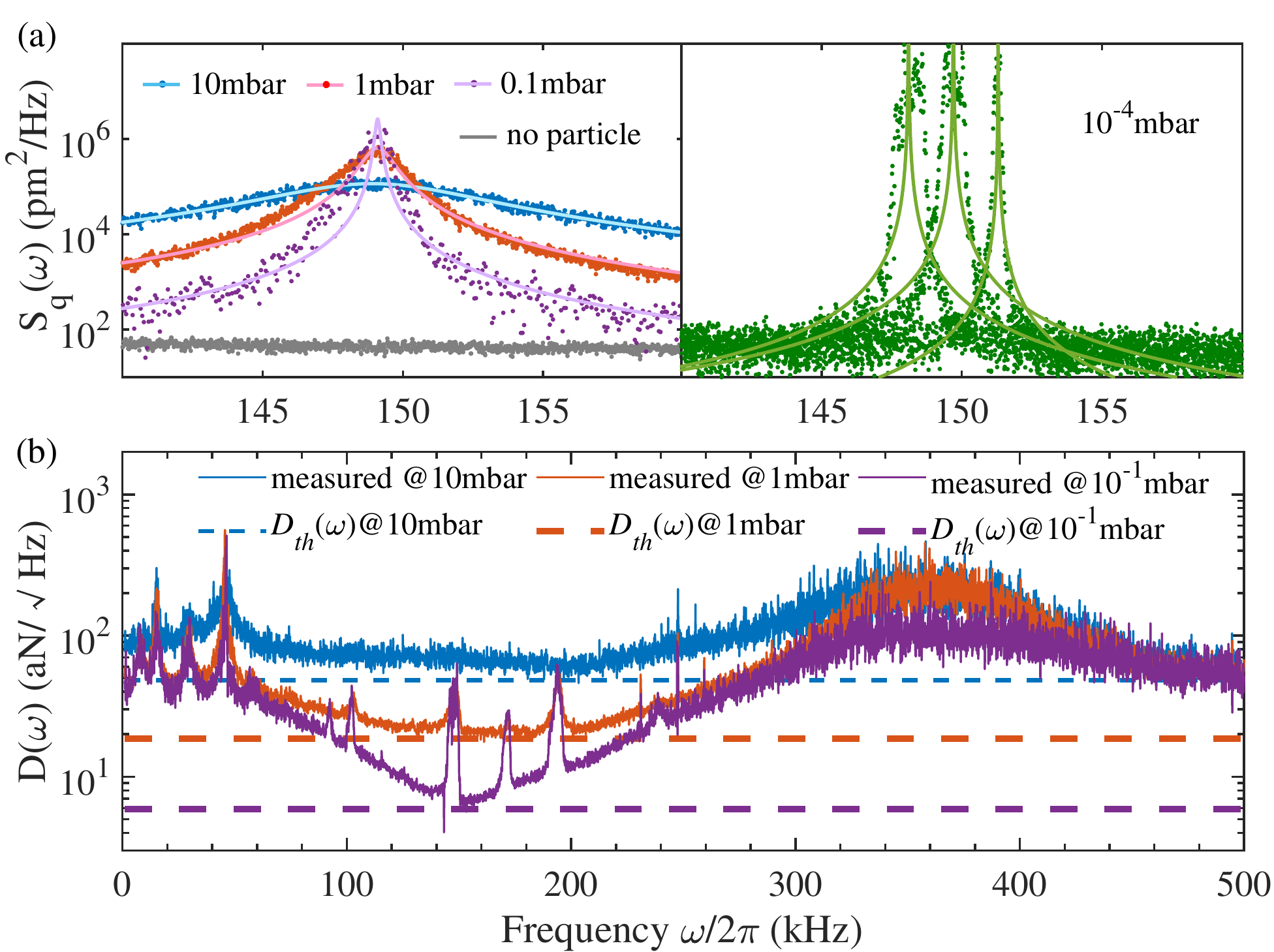}}
		\caption{ (a) PSD of  displacement and (b) FDSS under different pressures. The sampling rate was 937 kHz, and the sampling time was 270 ms. The curve at high pressure was obtained by averaging 10 sets of data. The three green curves at $ 10^{-4}\rm mbar $ indicate the instability of the resonant frequency.
		The dotted line represents the thermal noise limit force detection sensitivity deduced from Eq.\ref{det-sen}. 
		FFDS for $ 10^{-2}\rm mbar $ and lower pressure is not shown in the figure due to the lack of transfer function accuracy.}
		\label{FDSS}
	\end{figure}

	Employing the measured transfer function, we transformed the displacement response of the system absent of external electrical force into the FDSS according to Eq.\ref{eq-FDSS}, as shown in Fig.\ref{FDSS}. At pressures greater than 0.01 mbar, the PSD decreased with a decrease in pressure, as did the corresponding FDSS, which approached the theoretical thermal noise limit in the vicinity of the resonant frequency. 
	
	As we vacuumized further, thermal noise was not dominant in the system noise, which did not decrease with pressure. Although the transfer function still varied with pressure, the FDSS of the nanosphere was unable to decrease further. The actual sensitivity of the system was only of the order of $\rm10^{-17}N/Hz^{1/2}$ at frequencies far above or below the resonance, and the force detection performance of the oscillator faltered significantly owing to non-thermal noise from the light source, circuit system, and ambient vibration.
	
	An optimal detection sensitivity limited by thermal noise can be obtained near the resonance, which can still be improved by decreasing the pressure. However, a high vacuum also brings two significant challenges to the force detection of the resonator: contraction of the optimal detection frequency band and frequency instability induced by thermal nonlinearities \cite{2013Thermal}.
	At pressures below 1 mbar, the sharp resonant peak drifted greatly, correspondingly altering its electrical response signal. Therefore, accurately obtaining the transfer function value near the resonant frequency became relatively difficult, and a non-Lorentz PSD due to the broadening effect was unavoidable after multiple averaging, as shown in Fig\ref{FDSS}(a). Simultaneously, in an attempt to maintain stable suspension of the nanosphere, feedback cooling also crippled the actual electrical response signal, causing notable error in the measurement of the transfer function. Although feedback cooling did not change the force detection sensitivity of the system, accurately measuring the transfer function under the state of feedback cooling in a high vacuum remains an difficult obstacle, which we aim to tackle in the next stage of this research. 
	\section{Conclusions}
	In this paper, we proposed a calibration method based on the harmonic Coulomb force. We applied this method to calibrate the FDSS of a levitated nanomechanical sensor. The measured transfer function can be utilized to obtain the actual force detection sensitivity spectrum. In the linear range, the measured transfer function coincided with the theoretical curve. The calibration method is conducive to the translation of a levitated nanomechanical resonator into a practical force sensor. 
	
	We evaluated the current performance of the system in terms of FDSS, which more accurately describes the practical system performance compared with the thermal noise limited sensitivity. Under a high vacuum, the systemic force detection sensitivity was far from the fundamental limit  based on thermal noise. To improve the force detection performance of the system, it is particularly important to optimize other noise sources in systems. 
	
\begin{backmatter}
	\bmsection{Funding} National Natural Science Foundation of China (62005248, 62075193)
	\newline
	Major Scientific Research Project of Zhejiang Lab (2019MB0AD01, 2020MB0AA01).
	\bmsection{Disclosures} The authors declare no conflicts of interest.
\end{backmatter}

\noindent
	\bibliography{references}
	
	\bibliographyfullrefs{sample}

\end{document}